# An open access repository of images on plant health to enable the development of mobile disease diagnostics

David P. Hughes [1,2,3] & Marcel Salathé [4]

1 Department of Entomology, 2 Department of Biology, 3 Center for Infectious Disease Dynamics, Penn State University, University Park PA, USA

4 Digital Epidemiology Lab, School of Life Sciences, School of Computer and Communication Sciences, EPFL, Switzerland

**Corresponding authors**

Correspondence to David Hughes, dphughes@psu.edu and Marcel Salathé, marcel.salathe@epfl.ch

**Human society needs to increase food production by an estimated 70% by 2050 to feed an expected population size that is predicted to be over 9 billion people. Currently, infectious diseases reduce the potential yield by an average of 40% with many farmers in the developing world experiencing yield losses as high as 100%. The widespread distribution of smartphones among crop growers around the world with an expected 5 billion smartphones by 2020 offers the potential of turning the smartphone into a valuable tool for diverse communities growing food. One potential application is the development of mobile disease diagnostics through machine learning and crowdsourcing. Here we announce the release of over 50,000 expertly curated images on healthy and infected leaves of crops plants through the existing online platform PlantVillage. We describe both the data and the platform. These data are the beginning of an on-going, crowdsourcing effort to enable computer vision approaches to help solve the problem of yield losses in crop plants due to infectious diseases.**

## BACKGROUND & SUMMARY

**The origins and continued evolution of agriculture in the face of infectious diseases and pests**

Perhaps the greatest technological advance that humans have ever made has been the domestication of plants during the agricultural revolution 8-12,000 years ago at multiple sites around the world (Diamond 2002). These events created civilizations through the steady, predictable supply of calories that could be obtained with lower and lower amounts of energy expended through human work. The enormous increase in human population from around 1 billion in the early 1800s to our current situation of over 7.2 billion has been made possible by an efficient and productive agricultural basis (Borlaug 2000). However, this steady supply of calories is currently threatened, as many of the advances resulting from the Green Revolution of the 1950s are failing due to infectious diseases and pests spread by globalization and compounded by climate change (Bourne 2015).

Many histories have recorded the enormous impact that infectious diseases have had on growing food crops (Ainsworth 1981). The iconic example is the Irish Potato famine of 1845-49 where an overdependence on a single crop with little genetic diversity set the stage for a devastating decline in yield that resulted in 1.2 million deaths out of a population of 9 million (Woodham-Smith 1991). At the time, plant diseases were poorly studied and the Irish Potato famine is widely identified as the event that began the scientific discipline of plant disease (Carefoot and Sprott 1967). It also marks the formal system of extension where science based knowledge is provided to farmers by state employed extension workers (Jones and Garforth 2005). In the intervening 170 years, plant pathology has grown to become a robust discipline offering crop growers multiple solutions from chemical control to genetic engineering to integrated pest management. At the same time, formalized extension systems have likewise developed to provide knowledge to growers.

However, the fact that the global food supply is annually reduced by an average of 40% (Oerke 2006) demonstrates that our collective battle against diseases and pests of crop plants is not won. Indeed, the emergence and spread of novel and highly virulent crop diseases like the stem rust UG99 that attacks wheat, black pod in Cocoa and viral infections of Cassava suggest that the situation may in fact be worsening. This is troubling at a time when the UN FAO recommends we must in fact increase the food supply by 70% to feed the future population[1].

In high-income countries, where food supply is currently less of a problem, a cursory glance at the media shows an increasingly high level of concern about food production, and many people express unease at the dominance of "Big Ag", the increased focus on growing monocultures (e.g. corn) for both meat production and biofuels, as well as over GMOs (Bourne 2015). Since the 1980s we have witnessed an increase in smallholder farmers that seek to promote diverse practices and crops. The essence of such movements were captured in Michael Pollan's book, *The omnivore's dilemma: a natural history of four meal* (Pollan 2006). Recognizing this, public health institutions are increasingly encouraging consumers to grow their own food, as both the process of growing food (exercise through gardening) and the yield (fresh fruit and vegetables) contribute to beneficial public health outcomes. Indeed, gardening, and urban gardening in particular, has become increasingly popular in the past two decades, reversing a long trend of consumers becoming less and less involved in growing their food. Globally, Google Trends captures this and reports a steady increase of interest in "urban gardening"[2].

---

[1] http://www.fao.org/fileadmin/templates/wsfs/docs/expert_paper/How_to_Feed_the_World_in_2050.pdf
[2] https://www.google.ch/trends/explore#q=Urban%20gardening

Whereas many communities in high-income countries are choosing to grow foods at small scales, such small-scale farming is the norm in poorer parts of the world. This is an imposed necessity rather than a choice. In many countries (e.g. those in Sub-Saharan Africa) as much as 80% of the population are farmers with single families growing diverse crops on small (2-5 hectare) plots with minimal mechanical or chemical (fertilizer, pesticides) inputs (Sanchez 2015). At this scale the relative impact of yield gaps (the gap between potential and actual yield) is very high (Collier and Dercon 2009, Foley et al. 2011). A consistent wedge reducing our ability to close this gap are infectious diseases and pests. Since most subsistence agriculture today occurs around the tropics and since the biodiversity of all infectious diseases (of humans, animals and plants) is higher in the tropics then the pressure of diseases are greatest in these areas. It is commonplace for smallholder farmers to routinely lose 80-100% of a given crop to pests and diseases (Oerke 2006).

**PlantVillage: a tool for crop health**

Three years ago, we co-founded an online platform dedicated to crop health and crop diseases, called PlantVillage (available at www.plantvillage.org). This platform was modeled after popular online platforms in the computer programming domain, including Stack Overflow www.stackoverflow.com, a community driven forum where anyone can ask and answer questions related to programming. By providing answers that are being up-voted by the community, users can build an online reputation, captured by a numeric score. The higher the score, the more rights a user gets on the platform. For example, users need a certain score to be able to vote on other's contributions. An even higher score is needed in order to be able to edit other people's contributions. This model, which has worked very well in many different contexts, has also been successful in PlantVillage, and the platform has seen its traffic grow 250% year over year. In the fall of 2015, the platform welcomed the 2 millionth visitor to the site.

In addition to this crowdsourced problem-solving, we have also created a library of open access information on over 150 crops and over 1,800 diseases, accessible on the same website. This content has been written by plant pathology experts, reflecting information sourced from the scientific literature. However, as the site is targeted directly to food growers, rather the professional plant pathologists, great care has been taken to write the content in a way that is easy to understand. Currently, most content is written in English, but we have recently begun to translate it into French, Spanish, and Portuguese. We will continue to translate more content into more languages.

While human-assisted disease diagnosis is a powerful tool, we believe that the potential for machine-assisted disease diagnosis has enormous potential. Disease phenotyping, when done by humans, usually involves a visual analysis of the presentation of the disease on the plant. For some visual phenotypes, disease identification by visual cues is straightforward; for others, it may be more challenging. Nevertheless, the visual diagnosis, if possible, so far requires humans. However, despite the challenge of crop health on an increasingly crowded planet (food security), investment in training plant pathologists has not grown correspondingly, and often even decreased (Flood 2010).

If a visual diagnostic (by a human) is possible, then computational tools should, in principle, be able to support the human diagnostician. In many cases, a computational diagnostic tool would indeed be the only way to get a diagnosis, due to the absence of expert help in many parts of the world. Even where human diagnostic expertise is available, scaling it to match global demand is not trivial. Since the 1980's the UN FAO has promoted Farmer Field Schools which focused on improving crop health in developing world countries (Braun et al. 2000). More recently the Plantwise Clinic efforts of CABI have undertaken similar efforts (Nicholls 2015). While both are excellent, they are not scalable without the sort of investment seen in developed world countries 150 years ago (Jones and Garforth 2005). This implies that a computational system that could aid with disease diagnosis, either alone or as support, would be both enormously beneficial and inherently scalable if provided online. Such a system would need the ability to recognize a disease from an image, and would thus be an image recognition system based on artificial intelligence.

**Recent developments in software and leveraging power of groups**

In recent years, remarkable progress has been made towards the goal of developing artificial intelligence (Ghahramani 2015). Fueled by breakthroughs in machine learning algorithm development, cheap computing, and cheap storage of very large data sets, artificial intelligence has permeated our everyday digital experiences. Whether it is a product recommendation based on our past consumer history, the automatic detection of a friend's name based on an photo uploaded to a social media service, or the automatic language detection and translation of webpages - the accuracy of such tools (i.e. recommendation, visual recognition, translation) is now so high that consumers are adopting them very rapidly. Such services are not just at play in consumer arenas but are also being used in medical settings such high throughput screening of radiographs to detect signs of cancer (Wang and Summers 2012).

Machine learning is a computational way of detecting patterns in a given dataset in order to make inferences in another, similar dataset. A classical textbook example is the machine recognition of handwriting such as postal addresses on envelopes. In recent years, generic object recognition has made tremendous advances, and is now approaching human accuracy. In facial recognition for example, the DeepFace algorithm developed by Facebook researchers has achieved an accuracy that matches humans (Taigman et al. 2014). These developments have been fueled by a variety of advances, but the most striking breakthroughs have come from the field of neuronal networks, and convolutional neural networks in particular (Krizhevsky et al. 2012). Additionally, advances in computing chips and notably GPUs (Graphical Processing Units) which can be linked together to form networks of computers has also been a key innovation.

The development of these algorithmic breakthroughs has come from three main sources (LeCun et al. 2015). The first is traditional academic research at universities and other institutes of higher education, where machine learning and related fields are increasingly important domains not just within computer science, but many other fields as well. The second source is industry, in particular some of the key players of the digital economy such as Facebook, Amazon, and Google. Google research, for example, lists over 450 scholarly papers that it has published in Artificial Intelligence and Machine Learning alone [3]. Facebook and others are catching up quickly, especially as they manage to attract the world's top researchers, given their financial resources and privileged access to data sets. Finally, the third source is perhaps the most surprising - the crowd.

Crowdsourcing is an increasingly common practice of soliciting services from large groups of people online. It has been popularized by services such as Amazon Turk, where researchers and companies can ask large groups of people to provide some tasks or services in exchange for money (Paolacci et al. 2010). Traditionally, these crowdsourced tasks have been simple for humans to do (such as assessing the sentiment of short texts), but increasingly, crowdsourcing is used as a method to find solutions to very hard problems as well. One of the most famous recent examples was the 2009 Netflix prize, where Netflix offered $1 million to anyone who was able to improve their recommendation algorithm by 10%. It was awarded to a team of engineers who published the algorithm openly after the competition (in line with the competition rules). There are now numerous data science competition platforms, where researchers or companies can run competitions on their datasets, usually in exchange for monetary prices, or academic recognition. Another, more vertically focused example of crowdsourcing is ImageNet. The "ImageNet Large Scale Visual Recognition Challenge" has, in just 5 years, become the benchmark in visual object category classification. The challenge started in 2010 at Stanford University and has since attracted participants from over 50 institutions. Many of the key breakthroughs in visual image classification has come through participants in this challenge (Russakovsky et al. 2014).

---

[3] http://research.google.com/pubs/ArtificialIntelligenceandMachineLearning.html

In order to leverage the potential of crowdsourcing algorithmic development, we are releasing a data set of tens of thousands of images of healthy and diseased plants, labeled by plant pathology experts. Over 50,000 images of these are now stored on www.plantvillage.org, openly accessible and released under the Creative Commons Attribution-ShareAlike 3.0 Unported (CC BY-SA 3.0), with the clarification that algorithms trained on the data fall under the same license. We chose this license to ensure that any disease diagnostic algorithm developed on the data can be made freely available to anyone. The data set will continue to grow over the coming months and years. The list of crops is given in Table 1.

## METHODS

Examples of the images and different phenotypes are given in Figure 1. All the images in the PlantVillage database were taken at experimental research stations associated with Land Grant Universities in the USA (Penn State, Florida State, Cornell, and others). We are continuing to collect images, and in the future, the list of sources we draw from will increase. Experimental research stations (both public and private) offer the possibility of taking many images in a reduced amount of time. The majority of the images were taken by two technicians working as a team.

From field trials of crops infected with one disease, the technicians would collect leaves by removing them from the plant. The leaves were then placed against a paper sheet that provided a grey or black background. All images were taken outside under full light. The light could be strong sun or cloud and we intentionally sought a range of conditions as the end user (grower with a smartphone) will ultimately take images under a range of conditions. For each leaf, we typically took 4-7 images with a standard point and shoot camera using the automatic mode  is a standard digital camera (Sony DSC - Rx100/13  20.2 megapixels). The leaf was rotated around 360 degrees as we imaged. We found this was important as, depending on both the reflectance and the nature of the disease, multiple images allowed us to capture more data. For crops such as corn (*Zea mays*) and squash (*Cucurbita* spp.), the leaves were too large to capture in a single frame while retaining high resolution, close proximity views. In these cases, we took images of different sections of the same leaf. Once the images were collected, they were edited by cropping away much of the background and orientating all leaves so that they tip-pointed upwards.

## DATA RECORDS

The data records contain 54,309 images. The images span 14 crop species: Apple, Blueberry, Cherry, Corn, Grape, Orange, Peach, Bell Pepper, Potato, Raspberry, Soybean, Squash, Strawberry, Tomato. In containes images of 17 fungal diseases, 4 bacterial diseases, 2 mold (oomycete) diseases, 2 viral disease, and 1 disease caused by a mite. 12 crop species also have images of healthy leaves that are not visibly affected by a disease. Table 1 summarizes the dataset.

The data records are available through the website www.plantvillage.org. A file mapping each the URL of each image to the classification has been deposited at [insert link after crowdAI competition finishes].

|  | Fungi | Bacteria | Mold | Virus | Mite | Healthy |
|---|---|---|---|---|---|---|
| **Apple (3172)** | *Gymnosporangium juniperi-virginianae* (276) *Venturia insequalis* (630) *Botryospaeria obtuse* (621) |  |  |  |  | (1645) |
| **Blueberry (1502)** |  |  |  |  |  | (1502) |
| **Cherry (1906)** | *Podosphaera spp* (1052) |  |  |  |  | (854) |
| **Corn (3852)** | *Cercospora zeae-maydis* (513) *Puccinia sorghi* (1192) *Exserohilum turcicum* (985) |  |  |  |  | (1162) |
| **Grape (4063)** | *Guignardia bidwellii* (1180) *Phaeomoniella spp.* (1384) *Pseudocercospora vitis* (1076) |  |  |  |  | (423) |
| **Orange (5507)** |  | *Candidatus Liberibacter* (5507) |  |  |  |  |

| Crop | | | | | | |
|---|---|---|---|---|---|---|
| **Peach (2657)** | | *Xanthomonas campestris* (2291) | | | | (360) |
| **Bell Pepper (2475)** | | *Xanthomonas campestris* (997) | | | | (1478) |
| **Potato (2152)** | *Alternaria solani* (1000) | | *Phytophthora Infestans* (1000) | | | (152) |
| **Raspberry (371)** | | | | | | (371) |
| **Soybean (5090)** | | | | | | (5090) |
| **Squash (1835)** | *Erysiphe cichoracearum / Sphaerotheca fuliginea* (1835) | | | | | |
| **Strawberry (1565)** | *Diplocarpon earlianum* (1109) | | | | | (456) |
| **Tomato (18,162)** | *Alternaria solani* (1000) *Septoria lycopersici* (1771) *Corynespora cassiicola* (1404) *Fulvia fulva* (952) | *Xanthomonas campestris pv. Vesicatoria* (2127) | *Phytophthora Infestans* (1910) | Tomato Yello Leaf Curl Virus (5357) Tomato Mosaic Virus (373) | *Tetranychus urticae* (1676) | (1592) |

Table 1: List of crops and their disease status currently (April 4, 2016) in the PlantVillage database. Numbers in parentheses indicate the number of images of a particular class.

## TECHNICAL VALIDATION

We confirmed the identity of diseases by having the expert plant pathologists determine the disease states. These experts worked directly in the field with the two technicians providing the diagnosis. The states were determined based on standard phenotyping approaches used by plant pathologists. In many cases, the experts we worked with infected the crop directly using standard experimental approaches in plant pathology. In those cases the diagnosis was easy. In some cases, the diseases occurred in sentinel plots that the experimental research stations maintain in order to identify the presence of a disease in a given region. In those cases, the diagnosis was again done by the expert. We curate all images into the PlantVillage database using the diagnosis from the experts. Only expertly identified leaves are present in the database.

**Acknowledgements**

We are grateful to Yannis Jaquet for building and maintaining the image platform on PlantVillage. We are grateful to previous developers including Brian Lambert and Isaac Bromley. We are especially grateful to Dr Lindsay McMenemy for collating the knowledge inside the Plant Library on PlantVillage. We thank Kelsee Baranowski, Ryan Bringenberg and Megan Wilkerson for image collection. We thank for Kelsee Baranowksi for coordinating image gathering, curation and databasing. We thank Anna Sostarecz, Ashtyn Goodreau, Ethan Keller, Kaity Gonzalez, Kalley Veit, Parand Jalili for editing assistance. We thank the following for researchers for crop specific advice. Apple: Dr Kari Peters, Janet Robinson. Blueberry:Dr Kathleen Demchack. Citrus: Dr. Stephen Futch, Jamie Burrow. Corn: Randy Dreibelbis, Dr Iffa Gaffoor. Grape: Linda Weaver, Dr. Violeta Tsolova, Dr. James Obuya. Peach: Dr. Normal Lalancette Potato: Mike Peck, Dr. Xinshun Qu. Squash: Dr. Sasha Marine. Strawberry: Kathleen Demchak. Tomato: PSU: Dr Matt Sullenberger, Dr. Majid Foolad, Dr Beth Gugino, Sara May. GCREC:Dr. Sam Hutton, Leticia Kumar. NREC: Dr Matthews Paret. Cornell: Dr. Chris Smart, Zach Hansen. We are grateful to the Huck Institutes at Penn State University for the initial financial support trough the HITS Fund. We thank Peter Hudson, Andrew Read and Vivek Kapur for continued support. We are grateful to Eberly College of Science, College of Information Science, College of Engineering, College of Agriculture, Office of Outreach, Office of Research and Office of the Provost (all Penn State) for continuing support. We are also grateful to EPFL for financial support.

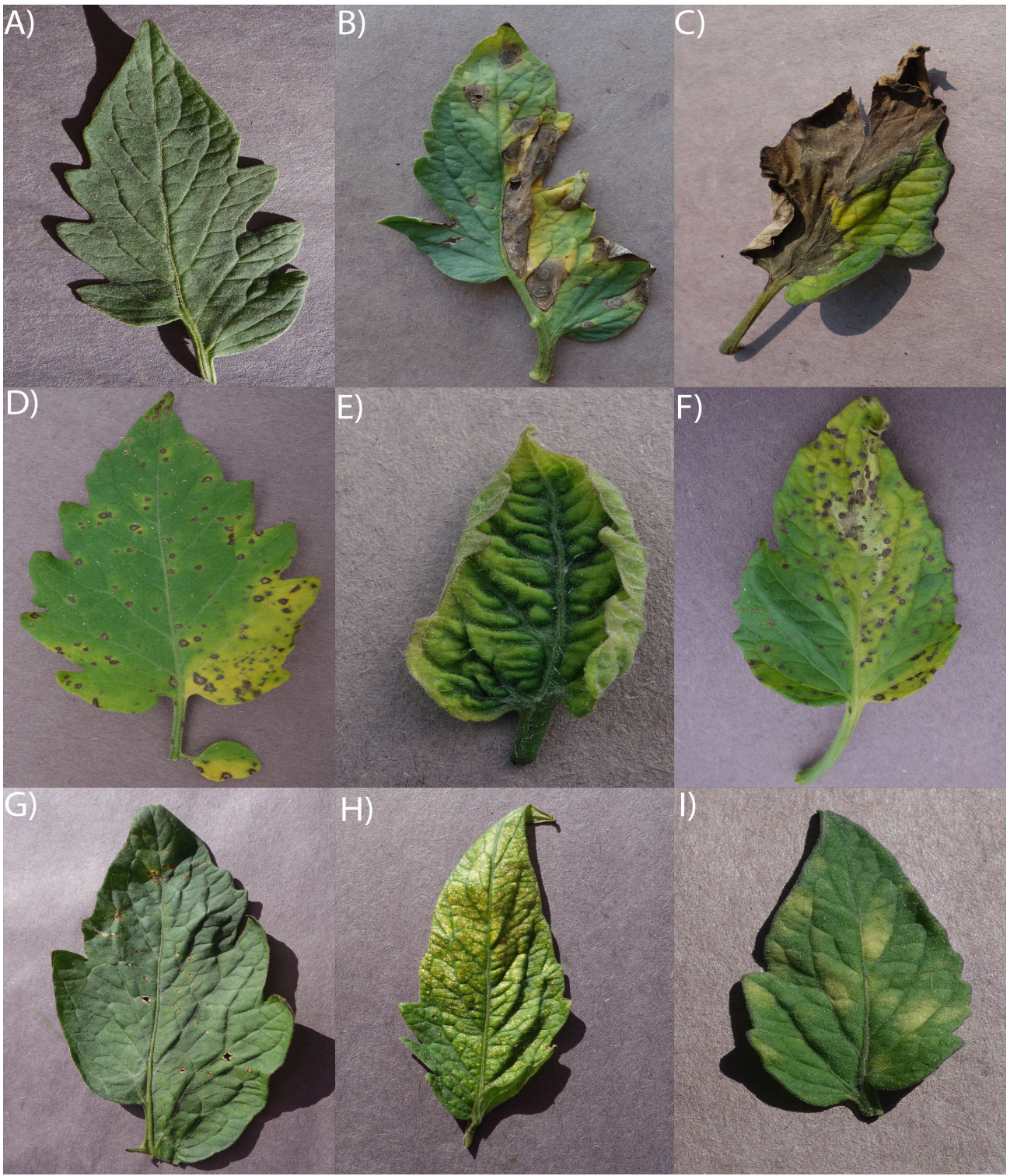

Figure 1: Examples of different phenotypes of tomato plants. *A)* Healthy leaf *B)* Early Blight *(Alternaria solani) C*) Late Blight *(Phytophthora Infestans) D)* Septoria Leaf Spot *(Septoria*

*lycopersici) E)* Yellow Leaf Curl Virus *(Family Geminiviridae* genus *Begomovirus) F)* Bacterial Spot *(Xanthomonas campestris pv. vesicatoria) G)* Target Spot *(Corynespora cassiicola) H)* Spider Mite *(Tetranychus urticae)* Damage